\documentstyle[prl,aps,epsf]{revtex}         

\begin{document}
%
 \twocolumn[\hsize\textwidth\columnwidth\hsize\csname
 @twocolumnfalse\endcsname

\title{Quantum tunneling across spin domains in a Bose-Einstein
condensate}
\author{D.M.~Stamper-Kurn, H.-J.~Miesner, A.P.~Chikkatur, S. Inouye,
J. Stenger, and W.~Ketterle}
\address{Department of Physics and Research Laboratory of
Electronics, \\
Massachusetts Institute of Technology, Cambridge, MA 02139}
\date{ } \maketitle

\begin{abstract}
Quantum tunneling was observed in the decay of metastable
spin domains in gaseous Bose-Einstein condensates.
A mean-field description of the tunneling was developed and compared
with measurement.
The tunneling rates are a sensitive probe of the boundary between
spin domains, and indicate a
spin structure in the boundary between spin domains which is
prohibited
in the bulk fluid.
These experiments were performed with optically trapped $F=1$ spinor
Bose-Einstein
condensates of sodium.
\end{abstract}
\pacs{PACS numbers:}
\vskip1pc
 ]

A metastable system trapped in a local minimum of the
free energy can decay to lower energy states in two ways.
Classically, the system may
decay by acquiring thermal energy
greater
than the depth of the local energy well (the
activation energy).
Yet, according to quantum mechanics, the system may decay even in the
absence of thermal fluctuations by tunneling through
the classically forbidden energy barrier.
Quantum tunneling describes a variety
of physical and chemical phenomena \cite{tunnelrefs,fowl28}
and finds common applications
in, for example, scanning tunneling microscopy.
In these systems, tunneling dominates over
thermal activation because the energy barriers are much larger than
the thermal energy.

Bose-Einstein condensates of dilute atomic gases
\cite{becrefs} offer a new
system to study quantum phenomena.
Recently, metastable Bose-Einstein condensates were
observed in which a configuration of phase-separated component
domains persisted for tens of seconds in spite of an external force which
favored
their rearrangement \cite{mies98meta}.
The metastability was due both to the restriction of motion to
one dimension by the narrow trapping potential
and also to the repulsive interaction between the domains.
Thermal relaxation to the ground state was identified and found to be
extremely slow, even at temperatures
($\sim$100
nK) much larger than the energy barriers responsible for
metastability ($\sim$5 nK), due to the scarcity of non-condensed
atoms,
to which the thermal energy is available.

In this article, we examine the decay of metastable spin
domains in an $F=1$ spinor condensate via
quantum tunneling.
The tunneling rates provide a sensitive probe of the
boundary between spin domains and of the penetration of
the condensate wavefunction into the classically forbidden region.
Tunneling barriers are formed not by an external
potential, but rather by the intrinsic repulsion between two immiscible
components of a quantum fluid.
These energy barriers are naturally of nanokelvin-scale height and
micron-scale
width in the presence of weak magnetic field gradients, and are thus
a promising tool for future studies of quantum tunneling and
Josephson oscillations
\cite{jose62,josebec,dalf96}.

\begin{figure}[ht]
\epsfxsize=70mm
 \centerline{\epsfbox{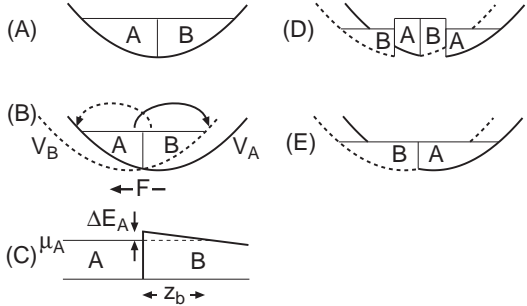}}
\caption{Metastable spin domains and the energy barrier for decay.
   (A) The ground state of a two-component condensate
   consists of two phase-separated domains.
   (B) A state-selective force $F$
   displaces the trap
   potential $V_{B}$ from $V_{A}$, creating metastable spin domains.
   Atoms tunnel from the metastable spin domains (direction of
   arrows) through an energy barrier (C) of
   maximum height $\Delta E_{A}$ and width $z_{b} \simeq \Delta E_{A} / F$
   (similar for component $B$).
   (D) Tunneling proceeds from the metastable domains (inner) to
   the ground-state domains (outer) until (E) the condensate has
   completely relaxed to the ground state.
}
   \label{scheme}
\end{figure}

We
begin by considering the one-dimensional motion of
a Bose-Einstein condensate
comprised of atoms of mass $m$ in two different internal states,
$|A\rangle$ and $|B\rangle$.   The condensate is held in a
harmonic trapping potential which has the same strength for each
component.
In a mean-field description, the
condensate wavefunction $\psi_{i}(z)$
is determined by two coupled
Gross-Pitaevskii equations \cite{ho96bin,esry97hf}
\begin{eqnarray}
\left( - \frac{\hbar^2}{2 m} \frac{d^{2}}{d z^{2}} + V_i(z) + g_i
n_i(z) +
g_{A,B} n_j(z) - \mu_i \right) \psi_i(z) & = & 0 \nonumber
\end{eqnarray}
where
$V_i(z)$ is the trapping potential, $n_{i}(z)$ the
 density and
$\mu_i$ the chemical potential of each component
($i,j = \{ A,B \}, i \neq j$).
The constants $g_A$, $g_B$, and $g_{A,B}$ (all assumed positive) are
given by $g = 4 \pi \hbar^2 a / m$
where $a$ is the $s$-wave scattering length which describes
collisions between atoms in the same ($a_A$ and $a_B$) or different
($a_{A,B}$) internal states.
Bulk properties of the condensate are well described by neglecting the
kinetic energy (Thomas-Fermi approximation). 
Under the condition $g_{A,B} > \sqrt{g_A g_B}$, the two components
tend to phase-separate (as observed in \cite{mies98meta,sten98spin}).
The ground state configuration consists
of one domain of each component on opposite sides of the trap
(Fig.~\ref{scheme}A).
The chemical potentials are determined by the densities
at the boundary $n_i^b$
as $\mu_i = g_i n_i^b$, and are
related to one another by the condition of equal pressure,
$\mu_A^2 / 2 g_A = \mu_B^2 / 2 g_B$.

Within the Thomas-Fermi approximation, the domain boundary is sharp
and the two components do not overlap.
Yet, the kinetic energy
allows each component to penetrate within the domain of the other.
The energy barrier for component $A$ (similar for $B$) is
$\Delta E_A(z) = V_{A}(z) + g_{A,B} n_{B}(z) - \mu_A$.
Neglecting slow variations in $V_A$ and $n_B$ gives the
barrier height
\begin{equation}
\Delta E_A = \mu_A \left( \frac{g_{A,B}}{\sqrt{g_A g_B}} - 1 \right).
\end{equation}

In this work we consider a condensate of atomic sodium in the two
hyperfine states $|A> = |F=1, m_{F} = 0>$ and $|B> = |F=1, m_{F} =
1>$, with scattering lengths of $a_{1} = a_{0,1} = 2.75$ nm
\cite{ties96} and $(a_{1} - a_{0}) = 0.10$ nm \cite{burk98}.
The barrier height for atoms in the $|m_{F} = 0\rangle$ state is then
$0.018 \, \mu_{0}$, a small fraction of the chemical potential.

Consider that a state-selective force $- F \hat{z}$ 
displaces the trapping potential $V_B(z)$ from
$V_{A}(z)$ (Fig.~\ref{scheme}B).
Due to the energy barrier discussed above, the atoms
cannot, classically, move to the other end of the
trap and thus the condensate is left in a high-energy configuration.
This configuration can decay by
tunneling.
At a domain boundary at the center at the condensate,
$d V_A / d z = g_B d n_B / dz = - F / 2$, and so
$\Delta E_A(z) = \Delta E_A -
F z (g_{B} + g_{A,B}) / 2 g_{B}$ (Fig~\ref{scheme}c).
The width of the barrier
becomes $z_b = \Delta E_A / F \times 2 g_B / (g_{A,B}+g_{B})$.
Tunneling from the metastable spin domains is analogous to the
field emission of electrons from cold metals \cite{fowl28}, where the
energy barrier height corresponds to the work function of the
metal and the force
arises from an applied electric field.
The tunneling rate $d{N}_{A}/ dt$ of atoms in state $|A\rangle$ from
the metastable spin domain is then given by the Fowler-Nordheim
relation \cite{fowl28}
\begin{eqnarray}
\frac{d N_A}{d t} & = &  \gamma \exp\left( - 2 \sqrt{\frac{2
m}{\hbar^2}}
\int_0^{z_b} \sqrt{\Delta E_A(z)} dz \right) \\
& = & \gamma \exp\left( -\frac{4}{3} \sqrt{\frac{2 m}{\hbar^2}}
\frac{2
g_B}{g_{A,B} + g_B} \frac{\Delta E_A^{3/2}}{F}\right)
\label{fowlnord}
\end{eqnarray}
where $\gamma$ is the total attempt rate for tunneling, and the
exponential is the tunneling probability.


The rate of quantum tunneling was studied in three steps.
First, condensates of sodium in the
$|F=1, m_{F} = -1\rangle$ hyperfine state were created
in a magnetic trap \cite{mewe96bec} and transferred to a
single-beam infrared optical trap \cite{stam98odt} with a $1/e^{2}$
beam radius of 12 $\mu$m, an aspect
ratio (axial / radial length) of about 60, and a depth of 1 -- 2
$\mu$K.
Chirped radio-frequency pulses were used to create two-component
condensates with nearly equal populations in the $|m_F = 0\rangle$ and $|m_F =
1\rangle$ states
\cite{mies98meta,sten98spin}.  Shortly afterwards, the two components
were
separated into two domains by the application of a
strong (several G/cm) magnetic field gradient along the axis of the trap
in a 15 G bias field.
The spin domains were typically 100 -- 200 $\mu$m long.

Second, the condensates were placed in a metastable state by
applying a magnetic field
gradient $B'$ in the opposite direction of that used to initially
separate the components \cite{mies98meta}.
This metastable state
corresponds to that shown in Fig.~\ref{scheme}B,\
where we identify the states $|A\rangle = |m_{F} = 0\rangle$ and
$|B\rangle =
|m_{F} = 1\rangle$.
The field
gradient exerted a state-selective force $F = g \mu_B m_F B'$ where
$g = 1/ 2$ is the Land{\'e} g-factor and $\mu_B$ the Bohr magneton.
The condensate was then allowed to evolve freely at the
gradient $B'$ and a bias field $B_0$ for a variable time $\tau$ of up
to 12 seconds.

Finally, the
condensate was probed by time-of-flight absorption
imaging combined with a Stern-Gerlach spin separation
\cite{mies98meta,sten98spin}.  The radial expansion of the condensate
in time-of-flight allowed for independent measurement of the
chemical potentials $\mu_0$ and $\mu_1$ \cite{mewe96bec}, while the
axial distribution allowed for measurement of
the number of
atoms in the metastable and ground-state domains of each spin
state.

\begin{figure}[ht]
\epsfxsize=65mm
 \centerline{\epsfbox{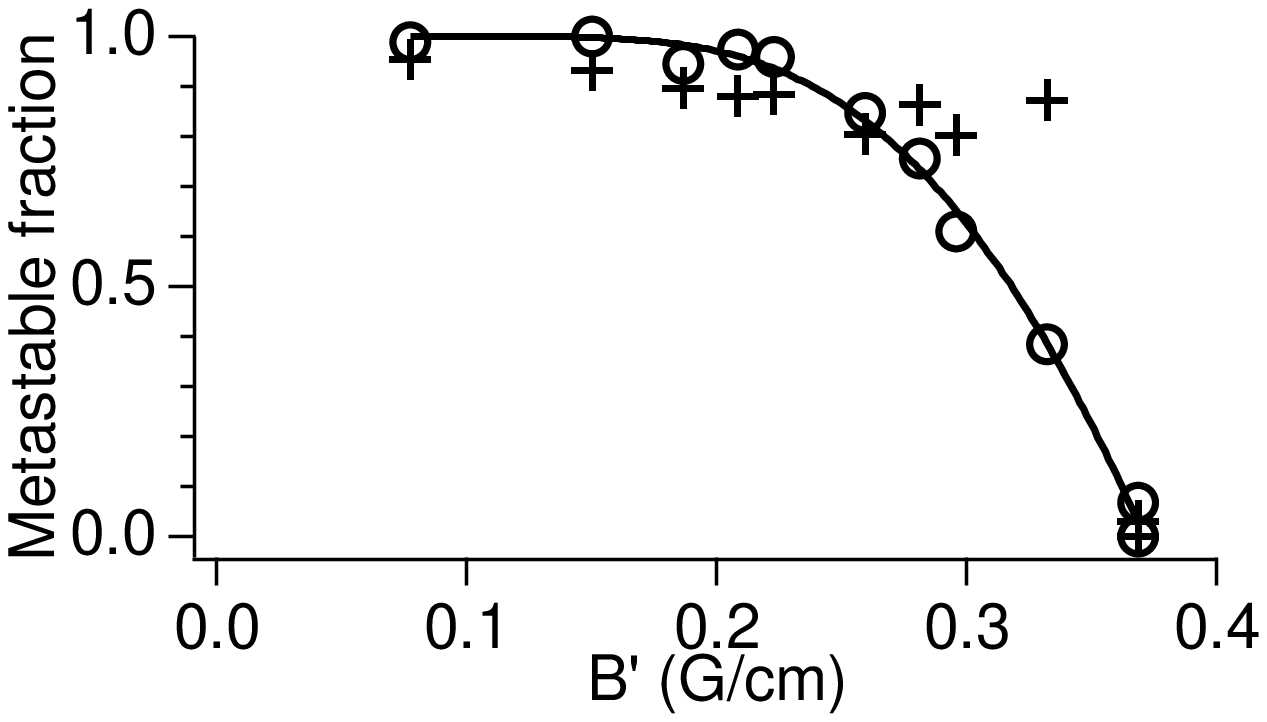}}
\caption{Tunneling across a barrier of variable width.
Condensates at constant density were
probed after 2 seconds of tunneling at a variable field gradient $B'$.
The fraction of atoms of each spin state in their metastable
domain is shown.
Circles represent the $m_{F} = 0$ atoms, and pluses the $m_{F} =
1$ atoms.
A fit to the $m_{F} = 0$ data (solid line)
determines the barrier attempt rate and tunneling probability.
The data indicate that the tunneling rate for $m_F=0$ atoms is larger
than that for $m_F=1$ atoms.}
   \label{gradscan}
\end{figure}

The mean-field description of tunneling from the metastable spin
domains was tested by measuring the tunneling rate across
energy barriers of constant height and variable width.
Condensates in a 15 G bias field with a chemical potential $\mu_{0} = 300$ nK
were probed after 2 seconds of tunneling at a variable field gradient
$B'$ (Fig.~\ref{gradscan}).
Thus, the energy barrier for tunneling had a constant height of 5 nK,
and a width between 4 and 20 $\mu$m.
As the barrier width was shortened by increasing $B'$,
the fraction of atoms in the $m_{F} = 0$ metastable
 spin domains decreased.
As expressed in Eq.~\ref{fowlnord}, the number of atoms which tunnel
 from the metastable to the ground state domains in a time 
$\tau$ should vary as
$\gamma \tau e^{- \alpha / B'}$
where $\gamma$ and $\alpha$ were
determined by fits to the data as $\gamma = 1.5(5) \times 10^{7} \,
\mbox{s}^{-1}$ and $\alpha = 1.5(2) \, \mbox{cm}/\mbox{G}$.
This value of $\alpha$ gives a tunneling probability of about $e^{-4}$
for  $B' = 370$ mG/cm, at which the metastable domains were fully depleted.

The tunneling attempt rate $\gamma$ can be estimated as the product of
two factors.
First, a bulk flux can be estimated by
considering the pressure $g_{0}
n_{0}^{2}/2$
to arise from an incoming atomic flux $n_{0} v/2$ which collides
elastically
at the boundary, imparting an impulse $2 m v$ per particle.  This
gives
$\gamma_{bulk} = \langle n_{0} v_{s} \rangle_{rad} / \sqrt{2}$
where $v_s = (g_{0} n_{0}/m)^{1/2}$ is the
Bogoliubov
speed of sound \cite{soundprop} and $\langle \ldots \rangle_{rad}$
denotes an integral over the radial dimension of the condensate.
This bulk flux is reduced an extinction factor $f$ which accounts
for the interpolation of the condensate wavefunction between the
bulk spin domain and the classically forbidden
region.
We use the treatment of Dalfovo {\it et al.}\
\cite{dalf96},
by considering the boundary between the two spin
domains not as a sharp division (as in Fig.~\ref{scheme}C), but
rather as a region of width
$d = (\hbar^{2} / 2 m \Delta E_{0})^{1/2} \simeq 1.5 \, \mu$m
wherein the density $n_{1}(z)$
rises linearly between $0 \leq z \leq d$.
We then find the density of the $m_{F} = 0$ component at the edge of the
boundary region $n_{0}(d)$
to be reduced from its bulk value by a
factor $f \simeq 1/10$.
Using $\mu_{0} = 300$ nK and a radial trap frequency of 500 Hz
 gives an estimate of $\gamma_{bulk} \simeq  10^{8} \, \mbox{s}^{-1}$ and
 $\gamma \simeq  10^{7} \, \mbox{s}^{-1}$.

The measured value of $\alpha$ can be compared with the prediction of
the Fowler-Nordheim equation (Eq.~\ref{fowlnord}).
Using the scattering lengths above and $\mu_{0} = 300$
nK gives $\alpha = 1.5(2) \, \mbox{cm}/\mbox{G}$,
in agreement with our
measurement (the error reflects a 10\% systematic uncertainty in
$\mu_{0}$).

In addition, $g_{1} > g_{0}$ implies $\mu_{1} >
\mu_{0}$ and thus the tunneling rate of $m_{F} = 1$ atoms
across the $m_{F} = 0$ domain should be {\it slower} than that of
the $m_{F} = 0$ atoms across the $m_{F} = 1$ domain.
The data in Fig.~\ref{gradscan} show evidence for this behavior.

\begin{figure}[ht]
\epsfxsize=65mm
 \centerline{\epsfbox{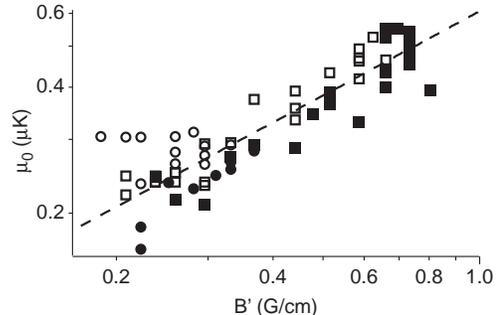}}
\caption{Threshold behavior for tunneling.
The chemical potential $\mu_{0}$ and gradient $B'$ are shown on a
logarithmic scale.  Closed symbols represent condensates which
had
fully decayed to the ground state, and open symbols those
which
had not.  Data were taken at two different settings of the optical
trap depth $U$ and tunneling time $\tau$:
$U = 1.0 \, \mu$K and $\tau = 2$ s (circles), and $U = 2.0 \, \mu$K
and $\tau = 1$ s (squares).
The dashed line shows a $\mu_0 \propto
B'^{\, 2/3}$ dependence for the $U = 2.0 \, \mu$K threshold.}
   \label{nbprime}
\end{figure}

The dependence of the tunneling rate on the energy barrier height was
probed by varying the condensate density.
For this, the number of trapped atoms was varied between about
$10^{5}$ and $10^{6}$ by allowing for a
variable duration of trap loss \cite{stam98odt} before creating the
metastable state.
Figure \ref{nbprime} shows data collected at two different settings of
the optical trap depth $U$ and tunneling time $\tau$ (see caption).
For each data series, at a given field gradient $B'$, there was a
threshold value of the chemical potential $\mu_{0}$
below which the condensates
had relaxed completely to the ground state, and above which they had
not.
Since the total condensate number and the attempt rate $\gamma$
should both scale as $\mu_{0}^{5/2}$ \cite{mewe96bec}, one
expects the threshold chemical potential for complete tunneling
to the ground state to vary as $\mu_{0} \propto B'^{\, 2/3}$.
The data shown in Figure \ref{nbprime} suggest a slightly steeper
dependence.

The chemical potential thresholds were approximately the same for both
settings of the optical trap depth.
Varying the optical trap depth $U$ also varied the temperature ($T
\simeq
1/10 \, U$ \cite{stam98odt}), and trap frequencies ($\omega \propto
U^{1/2}$).
That the threshold is independent of temperature confirms that
the decay proceeds by quantum tunneling rather than
thermal activation.
That the threshold is independent of the trap frequencies confirms
that the decay occurs by quantum tunneling of one spin component
through the other, rather than by radial motion of one component
around the other.

\begin{figure}[ht]
\epsfxsize=65mm
 \centerline{\epsfbox{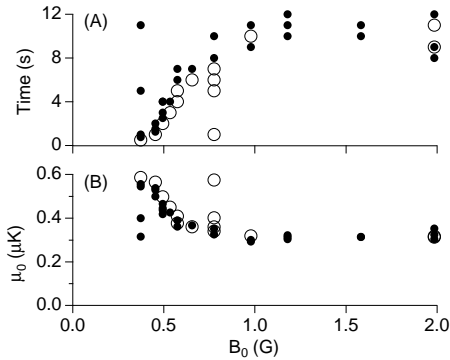}}
\caption{Variation of tunneling threshold with magnetic bias field
$B_{0}$.  Condensates probed after a variable tunneling time
 are represented by a closed symbol if total relaxation to the
ground state was observed, and with an open symbol if not.
As the field was lowered, condensates (A) relaxed in shorter times
and thus (B) at higher chemical potentials $\mu_{0}$.}
\label{bscan}
\end{figure}

Thus, we have shown the decay of the metastable spin domains at high
magnetic fields (15 G) to be
due to quantum
tunneling in a two-component condensate.
At lower magnetic fields, a dramatic change in the tunneling behavior
was observed.
Metastable spin domains of initial
chemical potential $\mu_{0} = 600$ nK
were prepared at a constant field gradient
of $B'= 130$ mG/cm, and a 15 G bias field.  The field was then ramped
down
to between 0.4 and 2 G within 10 ms, and held at a constant value
$B_{0}$.
After a variable tunneling time $\tau$ of up to 12 seconds, the
condensates were probed and evaluated as to whether they had
fully decayed to the ground state.
During the tunneling time, the chemical potential dropped
due to the loss of atoms from the trap.
As the field was lowered below about 1 G,
relaxation to the ground state occurred at earlier times
 (Fig.~\ref{bscan}A), and thus at
higher chemical potentials (Fig.~\ref{bscan}B).


The increase in the tunneling rates at lower magnetic fields
is inconsistent with the dynamics of a two-component condensate.
Our measurements thus serve as a probe of the spin domain
boundary and reveal the presence of the third $F=1$ spin component ($m_{F} =
-1$).
Atoms in the $|m_{F} = -1\rangle$ state are created by spin-relaxation
wherein two $m_{F} = 0$ atoms collide to produce an $m_{F} = 1$ and an
$m_{F} = -1$ atom
\cite{sten98spin}.
Due to the quadratic Zeeman effect,
the magnetic energy of two $m_{F} = 0$ atoms is lower than that of
their spin-relaxation product by
$2 q = 2 \times 20 \, \mbox{nK} \times (B_{0} / \mbox{G})^{2}$.
Interactions give rise to a spin-dependent energy term $c
\langle {\bf F} \rangle^{2}$, where
$c = \Delta g n / 2$, $\Delta g = g_{1} - g_{0}$, and $n$ is the
condensate density.

Neglecting the kinetic energy,
atoms in the $|m_{F} = -1\rangle$ state are excluded
from the domain boundary when $q > c/2$, i.e.\ at fields
$B_{0} \gtrsim 250$ mG for typical conditions ($n \sim 3 \times 10^{14} \,
\mbox{cm}^{-3}$) \cite{sten98spin}.
However, when the kinetic energy is considered, atoms in the
$|m_{F} = -1\rangle$ state are found to populate the boundary even at
fields $B_{0} > 250$ mG.
Within the boundary region, the average magnetization $\langle F_z \rangle$
must vary smoothly.
Minimizing the energy functional $q \langle F_{z}^{2}
\rangle + c \langle {\bf F} \rangle^{2}$ at constant magnetization
$\langle F_{z} \rangle$
indicates that,
for $q/2c \gg 1$, the fraction of atoms in the $|m_F =
-1\rangle$ state scales roughly as $B_{0}^{-2}$ \cite{calcfn}.
At a field of 1 G, the fraction of atoms in the domain boundary
in the $|m_{F} = -1\rangle$
state is at most $\sim 2$\% (about 300 atoms).

The presence of the $m_{F} = -1$ atoms in the barrier weakens the 
effective repulsion between the spin domains.
Consider a two-component system as before where $|A\rangle = |m_{F} =
0\rangle$ and  $|B\rangle =
\cos\theta \, |m_{F} =
1\rangle - \sin\theta \, |m_{F} = -1\rangle$ where $0 \leq \theta
\leq \pi/2$.
Evaluating the spin dependent interaction energy
\cite{ho98,sten98spin}
one finds $g_{B} = g_{0} + \Delta g \cos^{2}2
\theta$ and $g_{0,B} = g_{0}  + \Delta g (1 -
\sin 2 \theta)$.
Thus, as the fraction of atoms in the $|m_F=-1\rangle$ state is increased,
the repulsion of the $m_F=0$ atoms at the domain walls is weakened,
increasing the tunneling rate.


In conclusion, we have identified and studied quantum tunneling
across phase-separated spin domains in a Bose-Einstein
condensate.
The energy barriers due to the interatomic repulsion are a
small fraction of the chemical potential, and their width is simply
varied by the application of a weak force.
The tunneling rates at high field ($B_{0} > 1 G$)
were described by a mean-field model and an
application of the Fowler-Nordheim equation, while
the tunneling at lower fields reveals a change in the spin-state
composition of the domain boundaries.
Future studies using metastable spin domains as tunneling barriers
may may focus on the
roles of coherence and damping in quantum tunneling.
In the current setup, rapid Josephson oscillations might be expected at
frequencies ($\sim1$ kHz) given by the energy difference between the
metastable and ground state spin domains.
Over long time scales such oscillations
are presumably damped as the system evolves toward the ground state.
While no evidence for oscillatory behavior was found in the present work,
the use of smaller spin domains and better time resolution is
warranted.

This work was supported by the Office of Naval Research, NSF, Joint
Services Electronics Program (ARO), NASA, and the David and Lucile
Packard
Foundation.
A.P.C.\  acknowledges additional support from the NSF, D.M.S.-K.\
from JSEP, and J.S.\ from the Alexander von Humboldt-Foundation.


\begin{thebibliography}{10}
\bibitem{calcfn}
For $q/2c \gg 1$, the fraction of atoms in the $|m_F = -1\rangle$ state is $(1
  - \langle F_z \rangle)^2 \langle F_z \rangle / (q / 2c + (1 - 2 \langle F_z
  \rangle))$.
\end{thebibliography}

\end{document}